\documentclass[12pt]{iopart}
\usepackage{iopams,mathenv}
\usepackage{theorem}
\usepackage{graphicx,graphics}
\usepackage{amssymb,bbm}
\usepackage{euscript,pdfsync}

\newcommand{\lv}{\left \vert}

\newcommand{\ra}{\right \rangle}
\newcommand{\ket}[1]{\lv #1 \ra}

\theorembodyfont{\rmfamily}

\renewcommand{\vec}[1]{\boldsymbol{#1}}

\newcommand{\N}{\cal N}

\newcommand{\id}{\mathbbm{1}}

\newcommand{\gr}[1]{\boldsymbol{#1}}
\newcommand{\kebra}[2]{\vert #1 \rangle \! \langle #2 \vert}

\newcommand{\sig}{\gr{\sigma}}

\newcommand{\text}[1]{\textrm{#1}}
\newcommand{\eqref}[1]{(\ref{#1})}

\newcommand{\av}[1]{\langle{#1}\rangle}

\begin{document}

\title{Input-output Gaussian channels: theory and application}
\author{Tommaso Tufarelli,$^{1,2}$ Alex Retzker,$^{3,4}$ Martin B. Plenio,$^{3}$ Alessio Serafini,$^{2}$}

\address{$^1$ QOLS, Blackett Laboratory, Imperial College London, London SW7 2BW, UK}
\address{$^2$ Department of Physics \& Astronomy, University College London, 
Gower Street, London WC1E 6BT}
\address{$^3$ Institut f\"ur Theoretische Physik, Albert-Einstein Allee 11, Universit\"at Ulm, 89069 Ulm, Germany}
\address{$^4$ Racah Institute of Physics, The Hebrew University of Jerusalem, Jerusalem, 91904, Israel}
\begin{abstract}
Setting off from the classic input-output formalism, we develop a theoretical framework to characterise the Gaussian quantum channels 
relating the initial correlations of an open bosonic system to those of properly identified output modes. 
We then proceed to apply our formalism to the case of quantum harmonic oscillators, such as the motional degrees of freedom of trapped ions or nanomechanical oscillators, interacting with travelling electromagnetic modes through cavity fields and subject to external white noise. Thus, we determine the degree of squeezing that can be transferred from an intra-cavity oscillator to light, and also show 
that the intra-cavity squeezing can be transformed into distributed optical entanglement if one can access both output fields of a two-sided cavity.
\end{abstract}
\vspace{2pc}
\submitto{\NJP}

\maketitle
\section{Introduction}
Interfacing static and flying quantum degrees of freedom is a key step towards the realisation 
of operational quantum technologies.
In fact, such a requirement figures as one of the additional `networkability' Di Vincenzo criteria, 
central to quantum communication, distributed computation and any sort of quantum networking \cite{divincenzo}.
In current experimental set-ups, a promising approach to this problem consists in coupling a quantum degree of freedom with cavity light, and then to adopt the light leaking out of the cavity as the flying quantum degree of freedom 
\cite{parkins99,parkins00,Zhang03,morigi06,kimble08}. 
The latter may then be used to entangle distant trapped degrees of freedom by mixing on a beam-splitter \cite{martin} or achieve entanglement in a local site by detection of the emerging light \cite{martin2}. From the theoretical standpoint, this paradigm is well described by the seminal input-output formalism 
developed in the eighties by Yurke, Collett and Gardiner \cite{yurke84,collett84,wallsmilburn}.

In this paper, we consider general confined bosonic degrees of freedom (cavity light fields, trapped atoms, ions, optomechanical systems, 
or combinations thereof), adopt the input-output formalism to include interaction with a bath and a set of (possibly) accessible output modes,
and develop a very general theoretical framework to study the correlations of the output fields. 
Central to our analysis is the identification of suitable output modes, which are relevant to experimental detections in practical cases and 
will also lead us to the analytic determination of Gaussian channels \cite{demoen77, eisertplenio,eisert07} relating the initial correlations 
of the trapped system to the output correlations of the modes. 
If the initial state of the system is Gaussian, which is often the case when one considers bilinear Hamiltonians, the output fields will also be in a Gaussian state and the complete characterisation of their state will hence be provided in such a case.
Let us emphasise that the standard input-output approach would imply the numerical solution of rather convoluted integrations. 
Our treatment, instead, reduces the reconstruction of the output fields to an algebraic problem (involving only the computation of matrix inverses, matrix exponentials and the solution 
of Sylvester equations), which we completely characterise. Furthermore,  
such a treatment is wholly independent from the details of the quadratic intra-cavity dynamics 
and extends, up to a quadratic scaling in the computational resources, to any number of modes, even into regimes where brute-force integration 
would be impractical.

We then proceed to apply our theory to the relevant case of mechanical oscillators (such as trapped ions, particles, or vibrating mirrors in opto-mechanical set-ups) coupled to cavity light, and show that, given some initial degree of squeezing in the mechanical element, one can obtain squeezed travelling output light, or even distributed entanglement if one can access both output modes of a double-sided cavity. We also show how our formalism can provide a rigorous treatment of the continuous monitoring of mechanical motion, when performed via realistic (non-instantaneous) detectors and outside stationary regimes.
\section{General Formalism}
We consider a localised system comprised of $n$ bosonic modes, whose 
canonical operators are grouped together in the column vector $\hat{\vec{v}}(t)=({a}_1,{a}^{\dag}_1,\ldots,{a}_n,{a}^{\dag}_n)^\intercal$ and subject to the most general affine quantum Langevin equation:
\begin{equation}
\frac{{\rm d}\hat{\vec{v}}}{{\rm d}t} = A \hat{\vec{v}} + \hat{\vec{v}}_\text{\scriptsize \scriptsize in}(t) \; , \label{langio}
\end{equation}
where $A$ is the `drift matrix', determined by the system's own Hamiltonian and linear coupling to the environment (see below), 
and 
\begin{eqnarray}
    \!\!\!\!\!\hat{\vec{v}}_\text{\scriptsize in}(t) &=& K (a_\text{\scriptsize \scriptsize in,1}(t),a^{\dag}_\text{\scriptsize in,1}(t),\ldots,{a}_\text{\scriptsize in,n}(t),{a}^{\dag}_\text{\scriptsize in,n}(t))^\intercal + \bar{\vec{v}}_\text{\scriptsize in},\\
	K&=&{\rm diag}(\sqrt{\kappa_1},\sqrt{\kappa_1},\ldots,\sqrt{\kappa_n},\sqrt{\kappa_n}).\label{K-matrix}
\end{eqnarray}
Here, $\kappa_j$ quantifies the loss rate of the $j$-th mode of the system, which interacts with the input operators $\hat{a}_\text{\scriptsize in,j}$ and $a^{\dag}_\text{\scriptsize in,j}$, while $\bar{\vec{v}}_\text{\scriptsize in}\in{\mathbbm C}^{2n}$ is just a constant vector, satisfying $\bar{v}_{\text{\scriptsize in},2j-1}=\bar{v}_{\text{\scriptsize in},2j}^{*}$, 
accounting for the possible linear driving of the system. Given a quadratic Hamiltonian 
\begin{equation}
	\hat H=1/2\sum_{jk}H_{jk}\hat{v}_j\hat{v}_k+\sum_jV_j\hat{v}_j,
\end{equation}
with $H=H^\intercal$,$V_{2j-1}=V_{2j}^*$, $A$ and $\bar{\vec{v}}_\text{\scriptsize in}$ can be expressed as:
\begin{eqnarray}
	A&=&-i\Sigma H-K^2/2\;,\\
	\bar{\vec{v}}_\text{\scriptsize in}&=&-i\Sigma\vec V
\end{eqnarray}
where $\Sigma$ is the commutation matrix $\Sigma_{jk}=[\hat{v}_j,\hat{v}_k]$. Throughout the paper, we shall only consider white Gaussian input noise, characterised by $\langle\hat{\vec{v}}_\text{\scriptsize in}\rangle=\bar{\vec{v}}_\text{\scriptsize in}$ 
and by the following correlation functions:
\begin{eqnarray}
&&\!\!\!\!\!\!\!\!\langle \{\hat{v}_{\text{\scriptsize in},j}(t),\hat{v}^{\dag}_{\text{\scriptsize in},k}(t')\}\rangle - 2\bar{v}_{\text{\scriptsize in},j}\bar{v}_{\text{\scriptsize in},k}^{*}= ({\vec\sigma}_\text{\scriptsize in})_{jk}\delta(t-t'), \label{white}\\
&&\!\!\!\!\!\!\!\!\langle \hat{{v}}_{\text{\scriptsize in},j}(t) \hat{{v}}^\dagger_k(0)\rangle-\bar{{v}}_{\text{\scriptsize in},j}\langle\hat{{v}}^\dagger_k(0)\rangle=0,\label{uncorrelated}
\end{eqnarray}
where $\vec{\sigma}_\text{\scriptsize in}$ is related to a physical covariance matrix, as it obeys $K^{-1}\vec{\sigma}_{\text{\scriptsize in}}K^{-1}+i\Sigma\ge 0$. The output field is determined by the boundary condition: 
\begin{equation}
\hat{\vec{v}}_\text{\scriptsize out}=K\hat{\vec{v}}-K^{-1}\hat{\vec{v}}_\text{\scriptsize in} \; , \label{boundary}
\end{equation}
where we deviate slightly from the standard definitions (see, {\em e.g.}, \cite{wallsmilburn}) to keep the formulae that will follow more compact.
One can then insert the formal solution of Eq.~\eqref{langio} into Eq.~(\ref{boundary}) to obtain
\begin{equation}
\hat{\vec{v}}_\text{\scriptsize out}(t) = K {\rm e}^{At} \hat{\vec{v}}(0) + K \int_{0}^{t} {\rm e}^{A(t-s)} \hat{\vec{v}}_\text{\scriptsize in}(s) {\rm d}s - K^{-1}\hat{\vec{v}}_\text{\scriptsize in}(t) \, . \label{v}
\end{equation}
Note that, in general, not all the components of the output field $\hat{\vec{v}}_\text{\scriptsize out}$ will correspond to experimentally accessible modes. For example, the light leaking out of an optical cavity might be collected and further processed, 
while the `quantum information' that a mechanical mode dissipates into its phononic environment is lost for all practical purposes. In what follows, we shall assume that the accessible output fields do not suffer further losses before their experimental manipulation. However, as shown in \ref{two-sided}, the treatment of the lossy case simply amounts to combining the Gaussian channels described below with appropriate beam-splitter transformations.
\section{Exponential Pulses} \label{expo}
\begin{figure}
	\center{\includegraphics[width=.6\textwidth]{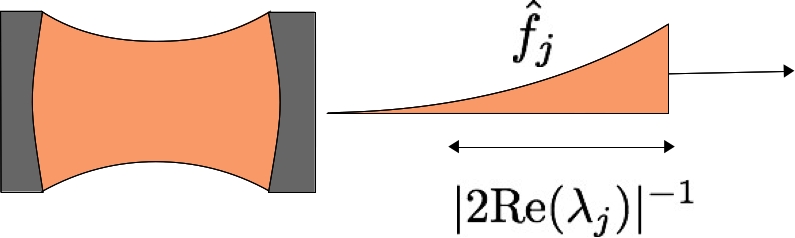}}
	\caption{The bosonic modes $\hat f_j$, as defined in Eq.~\eqref{f}, may represent travelling pulses of light emitted from a one-sided optical cavity. These pulses have an intensity profile that decays exponentially in time, with rate $|2{\rm Re(\lambda_j)}|$.  \label{figuraccia}}
\end{figure}

In many applications, one is interested in output modes characterized by an exponetial temporal profile (see Fig.~\ref{figuraccia} for an example). In general, this amounts to considering output modes of the form:
\begin{equation}
\hat{\vec{f}} = N \int_{0}^{\infty} {\rm e}^{\Lambda t} \hat{\vec{v}}_\text{\scriptsize out}(t) {\rm d}t \; , \label{f}
\end{equation}
where $\Lambda=\text{diag}(\lambda_1,\lambda_2,...,\lambda_{2n})=\text{diag}(\mu_1,\mu_1^*,...,\mu_n,\mu_n^*)$ is a $2n\times 2n$ diagonal matrix with ${\rm Re}(\Lambda)<0$, such that the mode profiles decay exponentially in time, with rates $|{\rm Re}\lambda_j|$, and no mixing between the output modes is allowed. $N=\left|2{\rm Re}(\Lambda)\right|^{1/2}$ is a (diagonal) normalisation matrix which guarantees that the $\hat{f}_j$ form a set of bosonic modes: $[\hat{f}_j,\hat{f}_k]=\Sigma_{jk}$.
The matrix integrations involved in deriving the explicit solution for the vector of modes $\hat{\vec{f}}$ are best dealt with by considering 
each component $\hat{f}_j$ separately: inserting Eq.~\eqref{v} into Eq.~\eqref{f} yields (see \ref{channelA}) 
\begin{equation}
\!\!\!\!\!\!\!\!\!\!\!\!\!\!\!\!\!\!\!\!\!\!\hat{f}_j=-\left[NK(A+\lambda_j \id)^{-1}\hat{\vec{v}}(0)+N(K(A+\lambda_j \id)^{-1}+K^{-1} )\int_{0}^{\infty}{\rm e}^{\lambda_j t} \hat{\vec{v}}_\text{\scriptsize in} {\rm d}t\right]_j \, , \label{fj}
\end{equation}
where $\id$ is the identity matrix in dimension $2n$. In this derivation, we have assumed that the system is stable, in the sense that $\lim_{t\rightarrow\infty}\exp(A t)=0$. The vector $\hat{\vec{f}}$ can be more compactly expressed by defining the 
elements of the matrix $X$ and the vector $\hat{\vec{u}}$ as per
\begin{eqnarray}
\!\!\!\!\!\!\!\!X_{jk}&=&\left[NK(A+\lambda_j \id)^{-1}\right]_{jk} \; , \label{X} \\
&&\nonumber\\
\!\!\!\!\!\!\!\!\hat{u}_j&=&\left[N(K(A+\lambda_j \id)^{-1}+K^{-1} )\int_{0}^{\infty}{\rm e}^{\lambda_j t} \hat{\vec{v}}_\text{\scriptsize in} {\rm d}t\right]_j .\label{uj}
\end{eqnarray}
Notice that the statistics of the vector $\hat{\vec{u}}$ are known, in that they can be reconstructed from Eq.~\eqref{white}.
In particular, one has:
\begin{equation}
Y_{jk}=\langle\{\hat{u}_j,\hat{u}^{\dag}_k\}\rangle-2\langle\hat{u}_j\rangle\langle\hat{u}^{\dag}_k\rangle = -\frac{\left(X'\vec{\sigma}_\text{\scriptsize in}{X'}^{\dag}\right)_{jk}}{\lambda_{j}+\lambda_k^*} \, , \label{Y}
\end{equation}
where $X'=X+NK^{-1}$. Then, if $\sig(0)$ is the symmetrised covariance matrix (CM) of the initial state of the system ($\sigma_{jk}(0)=\langle\{\hat{v}_j(0),\hat{v}^{\dag}_k(0)\}\rangle-2\langle\hat{v}_j(0)\rangle\langle\hat{v}^{\dag}_k(0)\rangle$), one can determine the output covariance matrix 
$\sig_{out}=\langle\{\hat{f}_j,\hat{f}^{\dag}_k\}\rangle-2\langle\hat{f}_j\rangle\langle\hat{f}^{\dag}_k\rangle$ from Eq.~\eqref{fj}, obtaining:
\begin{equation}
\sig_\text{\scriptsize out} = X \sig(0) X^{\dag} + Y \; . \label{ch1}
\end{equation}
Eq.~\eqref{ch1} defines a Gaussian quantum channel relating the system's CM at time $t=0$ to the output CM $\sig_\text{\scriptsize out}$ 
\cite{demoen77,eisert07}. 
Note that it was convenient for us to write the correlation matrices in terms of the field operators 
$a_{j}$ and $a^{\dag}_j$. In order to switch to a description in terms of quadratures, and retrieve the notation more commonly used in the recent 
literature \cite{eisert07}, one has simply to transform all the equations by similarity with the unitary \vspace{-.2cm}
$$U=\frac{1}{\sqrt{2}}\bigoplus_{j=1}^{n}\left(\begin{array}{cc}
1&1\\
-i&i 
\end{array}\right) \; .$$
\section{Slow detectors and time-dependent spectral measurements.}\label{spectro}
Let us now extend our treatment to the case of detectors with a finite response time, and to spectral analysers with finite resolution. In both cases, assuming a Lorentzian response in the frequency domain, the relevant output modes are of the form \cite{eberly}:
\begin{equation}
\hat{\vec{g}}(t) = N_t \int_{0}^{t} {\rm e}^{\Lambda (t-t')} \hat{\vec{v}}_\text{\scriptsize out}(t') {\rm d}t' \; , \label{g}
\end{equation}
where the diagonal matrix $\Lambda$ has the same form as above, and
$N_t=\left|2{\rm Re}\Lambda\right|^{1/2}[\id-{\rm e}^{2{\rm Re}(\Lambda) t}]^{-1/2}$ guarantees that the modes $\hat g_j(t)$ are bosonic. As anticipated, the modes of Eq.~\eqref{g} have two main interpretations. On one hand, they may describe continuous detection of the output fields with `slow' detectors. In this case, $|2{\rm Re}(\lambda_j)|$ quantifies the bandwidth (inverse of the response time) of the $j-$th detector. On the other hand, they may be used to model spectral measurements with finite resolution, so that $|2{\rm Re}(\lambda_j)|$ quantifies the spectral resolution of the $j-$th frequency filter.
\begin{figure}
	\center{\includegraphics[width=.8\textwidth]{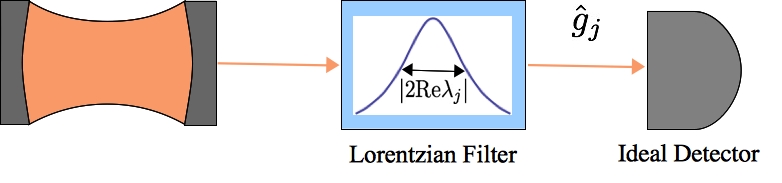}}
	\caption{Schematic representation for the modes $\hat g_j$. The output field of a cavity passes through a Lorentzian frequency filter [hence the convolution with an exponential in the time domain --- see Eq.~\eqref{g}], before reaching an ideal instantaneous detector. The quantity $|2{\rm Re}(\lambda_j)|$ may represent either the bandwidth of a realistic detector, or the frequency resolution of a spectral analyzer.
	\label{figuraccia2}}
\end{figure}

The exact treatment of this case, while more convoluted, runs along the same lines as the previous one, and the Gaussian channel can still be obtained in closed form as (see \ref{channelB})
\begin{equation}
\sig_\text{\scriptsize out} = Z \sig(0) Z^{\dag} + T \; , \label{ch2}
\end{equation}
with 
\begin{eqnarray}
Z_{jk} &=& \left[NK(A-\lambda_j \id)^{-1}\left({\rm e}^{At}-\id{\rm e}^{\lambda_jt}\right)\right]_{jk} \; , \label{Z} \\
&&\nonumber\\
T &=& FIF^{\dag} - G + H + (J+J^{\dag}) - (L+L^{\dag})  \; \label{T}, 
\end{eqnarray}
and
\begin{eqnarray}
F_{jk} &=& \left[NK(A-\lambda_j \id)^{-1}\right]_{jk} \; , \label{F} \\
&&\nonumber\\
G_{jk} &=& \frac{\left[(F+NK^{-1})\vec{\sigma}_\text{\scriptsize in}(F+NK^{-1})^{\dag}\right]_{jk}}{\lambda_j+\lambda_k^{*}}  \; , \label{G} \\ 
&&\nonumber\\
H_{jk} &=& {{\rm e}^{(\lambda_j+\lambda_k^{*})t}}G_{jk}  \; , \label{H} \\ 
&&\nonumber\\
J_{jk} &=& \left[F(A+\lambda_k^{*}\id)^{-1}\vec{\sigma}_\text{\scriptsize in}(F+NK^{-1})^{\dag}\right]_{jk}  \; , \label{J} \\ 
&&\nonumber\\
L_{jk} &=& \left[F{\rm e}^{(A+\lambda_k^{*}\id)t}(A+\lambda_k^{*}\id)^{-1}\vec{\sigma}_\text{\scriptsize in}(F+NK^{-1})^{\dag}\right]_{jk}  \, , \label{L} \\ 
&&\nonumber\\
I &=& \int_{0}^{t} {\rm e}^{At'}\vec{\sigma}_\text{\scriptsize in}{\rm e}^{A^{\dag}t'}  {\rm d}t' \label{I} \; .
\end{eqnarray}
Notice that even though the matrix $I$ cannot be given in a simple explicit form, 
it is the solution of the time-dependent Sylvester equation $AI+IA^{\dag}={\rm e}^{At}\vec{\sigma}_\text{\scriptsize in}{\rm e}^{A^{\dag}t}-\vec{\sigma}_\text{\scriptsize in}$,
and can be solved by standard linear algebra methods \cite{Sylvester}.

Clearly, the matrices $Z$ and $T$ defining the Gaussian channel are now time-dependent, as one should expect since the 
monitored output modes themselves depend on time in this approach.
\subsection{Stationary spectrum.}
Let us briefly mention that, 
if one assumes the stability condition $\lim_{t\rightarrow\infty}\exp(At)=0$, as well as ${\rm Re}(\Lambda)<0$,
the output field admits a stationary limit reflecting the status of the asymptotic steady state inside the cavity, 
with CM:
\begin{equation}
\sig_\text{\scriptsize out} =FIF^\dag - G  + (J+J^{\dag}) \; ,
\end{equation}
where all quantities are intended in the limit $t\to\infty$.
\section{Applications.}
Let us now apply our general formalism to instances of practical interest. 
We will deal with mechanical oscillators interacting with
cavity fields and, typically, concern ourselves with regimes 
where the cavity field can be adiabatically eliminated, yielding a direct coupling of the mechanical degree of freedom to the input and output fields of the cavity \cite{parkins99}. 
Exploiting this mechanism has become a standard practice to achieve cavity cooling of a mechanical oscillator, and to perform continuous detections of fluctuations in the oscillator's position \cite{optomechanics,machnes12,wilson}.
Here, we consider a further application which is readily available in the same setting. We are going to consider the possibility of high fidelity transfer of the quantum state of the motional degree of freedom, typically difficult to access, to a travelling pulse of light emitted by the cavity, which can be processed via standard optical tools (this process is sometimes referred to as `phonon-photon conversion'). In particular, we shall focus on the transfer of squeezing between matter and light and on the creation of distributed Gaussian entanglement between two output modes, given initial mechanical squeezing. In addition, we show how our formalism provides a rigorous and quantitative description of the continuous detection of the mechanical motion, performed via a realistic detector and in a manifestly non-stationary regime.

We investigate in detail the case in which the oscillator is provided by the motion of a trapped ion, keeping in mind that very similar results hold for optomechanical systems [in which case, Eq.~\eqref{coupling} below can be taken as the starting point]. Coupled ion-cavity systems have been realized experimentally in \cite{ionlight}, and cavity cooling of a trapped ion was proposed in \cite{zippilli05}.

Let us consider an ion with motional trapping frequency $\nu$ along a chosen axis, corresponding to the mode $a_1$. \footnote{We shall assume that the motion along the remaining axes is either negligible, or it can be factored out.} The ion is trapped inside an optical cavity that sustains a mode $a_2$ of frequency $\omega_c$, in turn coupled via a Jaynes-Cummings interaction of strength $g_0$ to an internal transition of the ion, described by a mode $a_3$ 
of frequency $\omega_0$. In order to couple the ion motion to the cavity field, we drive the ion internal transition with a detuned classical laser with Rabi frequency $\Omega$. In a frame rotating with the laser frequency $\omega_\textrm{\scriptsize L}$, operating in the Lamb-Dicke regime, and considering only quadratic terms in the bosonic operators \footnote{We recall that linear terms in the Hamiltonian can be accounter for by adding a constant to the vector $\bar{\vec{v}}_\text{\scriptsize in}$.}, we have the Hamiltonian (see \ref{ion-cavity})
\begin{eqnarray}
\hat{H} &= & \nu a_1^{\dag}a_1 + \delta a_2^{\dag}a_2 + \Delta a_3^{\dag}a_3 + 
g_0(a_2^{\dag}a_3+a_2a_3^{\dag})  
-\eta\Omega (a_1+a_1^{\dag}) (a_3+a_3^{\dag}) \; , \label{coupling2} 
\end{eqnarray}
where $\delta=(\omega_c-\omega_\textrm{\scriptsize L})$, $\Delta=(\omega_0-\omega_\textrm{\scriptsize L})$ and $\eta$ is the Lamb-Dicke parameter. As we shall see shortly, we are only interested in inducing virtual transitions of the ion, which will remain in its internal ground state with good approximation. Therefore, we only introduce a small error if we treat the mode $a_3$ as bosonic, which allows us to include the effects of spontaneous emission in the two-level system, all while remaining within the scope of our formalism and yielding an input-output Gaussian channel. 
In all our studies, we have simply verified a posteriori that the population of excited states above the first stays negligible at all times during the dynamics. As outlined in our general theory, we include loss rates in the problem via the matrix $K$ [see Eq.~\eqref{K-matrix}, where now $n=3$], where $\kappa_1$ is the motional heating rate per phonon, $\kappa_2$ the cavity loss rate and $\kappa_3$ the spontaneous emission rate of the ion's internal transition. The thermal excitations of each environment are instead included in the matrix $\vec{\sigma}_{\text{\scriptsize in}}$ [see Eq.~\eqref{white}]. In particular, only the phononic environment is thermally excited in our problem, and we shall indicate its thermal occupancy by $\bar n_\textrm{\scriptsize th}$. Note also that in this system only the output mode $a_{\textrm{\scriptsize out,2}}$ is accessible experimentally. A direct coupling between ion motion and cavity field can be achieved if the internal levels of the ion are eliminated by time-averaging the Hamiltonian \eqref{coupling2} in the large detuning regime $\Delta\gg(\nu,\delta,g_0,\Omega)$ \cite{james10,james11}. If the ion is initially in its internal ground state, one obtains the Hamiltonian 
\begin{equation}
\hat{H}' = \nu a_1^{\dag}a_1 + \delta' a_2^{\dag}a_2 + J(a_1+a_1^{\dag})(a_2+a_2^{\dag}) \; , \label{coupling}
\end{equation}
where $\delta'=\delta-g_0^2/\Delta$ and $J=\eta g_0\Omega/\Delta$. As anticipated, a Hamiltonian of this form is also readily available in optomechanical set-ups, 
where it provides a reliable description of the dynamics 
under strong driving of the cavity field \cite{optomechanics}.

The direct coupling between the oscillator and the cavity input-output fields is finally obtained by choosing $\delta'=\nu$, and considering the limit $\nu\gg\kappa_2\gg J\gg \bar n_\textrm{\scriptsize th}\kappa_1$. Then, neglecting the counter-rotating terms in Eq.~\eqref{coupling}, along with the adiabatic elimination of $a_2$ (see \ref{adiabatic}), yields the following input-output equations for the mechanical mode:\vspace{-.1cm}
\begin{eqnarray}
	&&\frac{{\rm d}a_1}{{\rm d}t}\simeq-i\nu a_1-\frac{\kappa}{2}a_1-i\sqrt{\kappa}a_{\textrm{\scriptsize in,2}}\;,\label{langio2}\\
	&&a_{\textrm{\scriptsize out,2}}\simeq-i\sqrt{\kappa}a_1+a_{\textrm{\scriptsize in,2}}\;\label{boundary2},
\end{eqnarray}
where $\kappa=4J^2/\kappa_2$ and we have also neglected the effects of the phononic environment ($\bar n_\textrm{\scriptsize th}\kappa_1\ll\kappa$ has to be checked a posteriori). As the photonic environment corresponding to $a_{\textrm{\scriptsize in,2}}$ is effectively at zero temperature, Eq.~\eqref{langio2} describes red-sideband cooling of the mechanical mode, with cooling rate $\kappa$. Moreover, the boundary condition \eqref{boundary2} provides us with the desired link between the cavity output field (the only detectable mode) and the motional degree of freedom. 

While a useful guidance, the two relationships (\ref{langio2}) and (\ref{boundary2}) are only approximations, whose reliability does not extend to the full dynamics we will be interested to study. 
However, the general analysis of Sections \ref{expo} and \ref{spectro} allows us to identify and treat exactly output modes that are suitable to study the applications we are interested in, without the need to resort to any approximation. 
We shall hence apply our general theory to those output modes, construct the input-output channels defined in Eqs.~\eqref{ch1} and \eqref{ch2}, 
and obtain exact quantitative results which are reliable even when the approximations 
leading to Eqs.~\eqref{langio2} and \eqref{boundary2} are only partially justified. 
In what follows, we shall consider an ion-cavity system characterised by typical experimental parameters, which we summarise in Table \ref{table}.

\begin{table}[t!]\begin{center}
\begin{tabular}{|l|c|}
\hline
Motional frequency of the ion: & $\nu=2\pi\cdot 5 {\rm MHz}$ \\ 
\hline
Jaynes-Cummings coupling strength: & $g_0=2\pi\cdot 0.62 {\rm MHz}$ \\ 
\hline
Laser Rabi frequency: & $\Omega=2\pi\cdot 1{\rm MHz}$ \\
\hline
Laser-internal transition detuning: & $\Delta=2\pi\cdot 10  {\rm MHz}$ \\
\hline
Lamb-Dicke parameter: & $\eta=0.08$\\
\hline
Cavity decay rate: & $\kappa_2=2\pi\cdot 53 {\rm kHz}$ \\
\hline
Spontaneous emission rate: & $\kappa_3=2\pi\cdot 360 {\rm kHz}$ \\
\hline
Phononic heating rate & $\bar n_\textrm{\scriptsize th}\kappa_1=2\pi\cdot24{\rm Hz}$  \\
\hline
Effective cooling rate & $\kappa\simeq2\pi\cdot 19 {\rm kHz}$\\
\hline
Detector's bandwidth: & $\Gamma= 2\pi\cdot 50 {\rm MHz}$ \\
\hline
\end{tabular}
\caption{Adopted values of dynamical parameters. These are obtained by considering the $230{\rm nm}$ transition of an Indium ion \cite{indium}, trapped at the center of a Fabry-Perot cavity of length $L\simeq1{\rm cm}$, waist $w\simeq 6 {\rm\mu m}$ and average photon lifetime $(\kappa_2)^{-1}=3\rm{\mu s}$. We have considered a stabilized trap with average heating time $(\bar n_\textrm{\scriptsize th}\kappa_1)^{-1}=6.6{\rm ms}$ at room temperature ($T=300$K). The effective cooling rate shown above corresponds to a temporal length $(\kappa)^{-1}\simeq 83 {\rm \mu s}$ for the mode $\hat f$. The $50$MHz bandwidth of the detector corresponds to an average response time $\Gamma^{-1}\simeq3.2{\rm ns}$. For further details, see \ref{experimental-parameters}.
\label{table}}\end{center}
\end{table}


\begin{figure}[t!]
\begin{center}
\includegraphics[width=.48\linewidth]{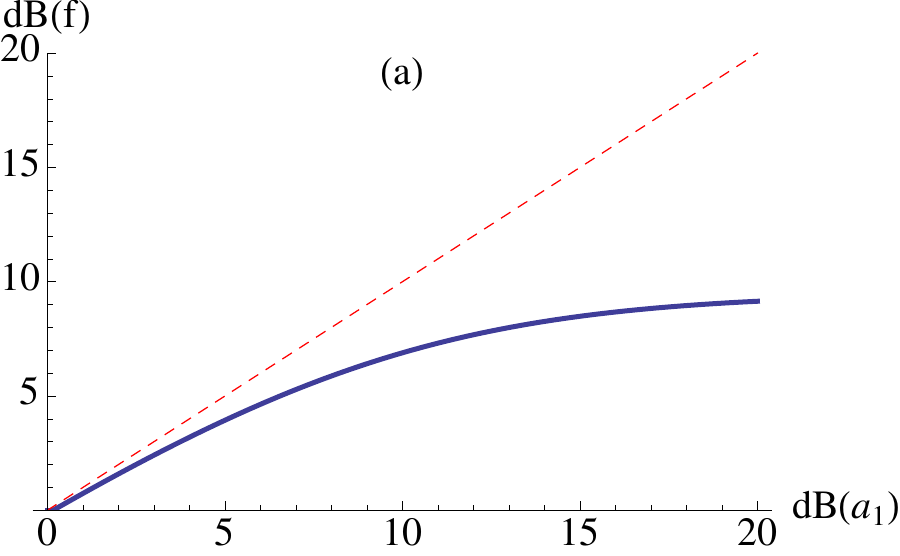}\includegraphics[width=.48\linewidth]{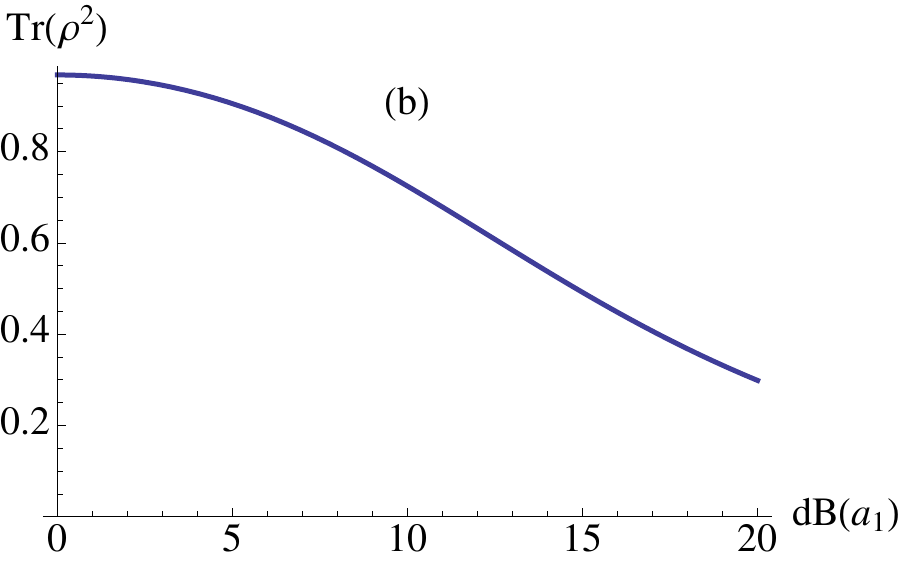}
\vspace{.2cm}\\
\includegraphics[width=.48\linewidth]{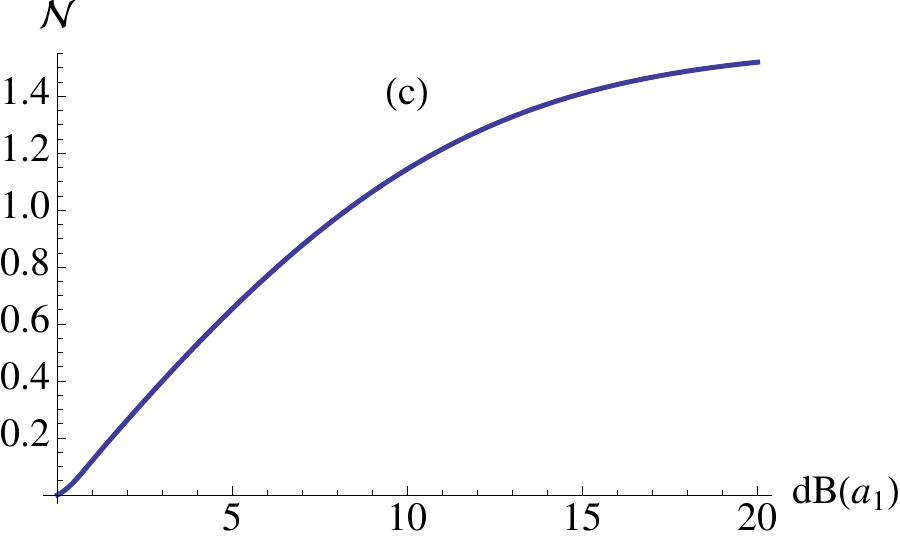}\includegraphics[width=.48\linewidth]{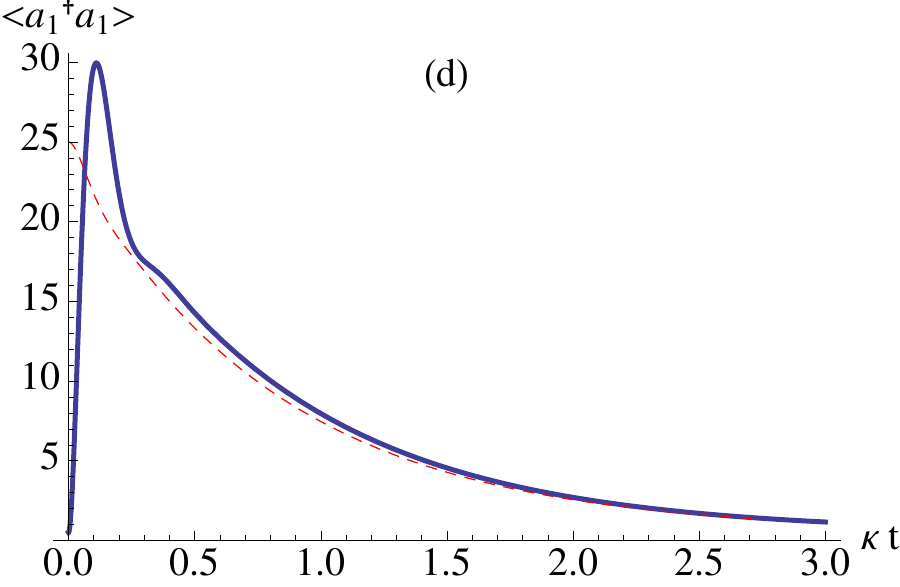}
\end{center}
\caption{(a): Squeezing, in ${\rm dB}$, of the output mode $f$ defined in Eq.~\eqref{exp-pulse}, as a function of the squeezing of the initial state of the motional mode $a_1$ (continuous blue curve). We have taken an initial pure Gaussian state, squeezed with respect to the quadrature $\hat{q}=(a_1+a_1^\dagger)$. For comparison, the straight line corresponding to the ideal case of perfect state transfer is also shown (dashed red curve). (b): Purity of the output state, quantified by $({\rm det}\vec\sigma_\textrm{\scriptsize out})^{-1/2}$ \cite{serafozzi03}, as a function of initial squeezing of the trapped ion. (c): Logarithmic negativity, in ebits, of the output modes leaking from a double-sided cavity, as a function of the initial squeezing of the trapped ion (we have assumed that the two mirrors have the same transmittivity). (d): Comparison between the expectation value of the motional excitation number $\langle a_1^\dagger a_1\rangle$ as inferred via Eq.~\eqref{temp-estimate} using a $50{\rm MHz}$-bandwidth detector (continuous blue curve), and the actual value (dashed red curve), as functions of rescaled time. After an initial transient, the two curves show excellent agreement. In all plots we have used the experimental parameters of Table~\ref{table}. \label{figura}}
\end{figure}

\subsection{Transfer of squeezing between matter and light}\label{squeeze-transfer}
Let us start by considering the possibility of transferring the initial state of the mechanical oscillator to a travelling pulse of light. This follows from standard input-output theory being applied to Eqs.~\eqref{langio2} and \eqref{boundary2}, yielding \cite{wiseman}
\begin{equation}
	f=\sqrt{\kappa}\int_0^\infty{\rm d}t\,{\rm e}^{(i\nu-\kappa/2)t}a_{\textrm{\scriptsize out},2}(t)\simeq-ia_1(0)\;,\label{exp-pulse}
\end{equation}
which means that the cavity emits a pulse of effective duration $\tau\sim1/\kappa$, whose quantum state is approximately the same as the oscillator initial state. Notice that, as we shall verify in the following, Eq.~(\ref{exp-pulse}) is only an approximation that will serve us as a guideline in the choice 
of the output modes, which we will however treat exactly through our general framework. 
In our language, the first equality in Eq.~\eqref{exp-pulse} corresponds to modes of the form \eqref{f}, with $\mu_2=i\nu-\kappa/2$. Note that the parameters $\mu_1,\mu_3$ can be set to any value, as we are only interested in the output field $a_{\textrm{\scriptsize out},2}$. In particular, we shall focus here on the case in which the mode $a_1$ is initially in a pure squeezed state, and investigate quantitatively the amount of squeezing transferred to the travelling mode $f$. The reliable realisation of such squeezing transfer constitutes a basic requirement for continuous variable quantum 
communication and information processing. Moreover, in the medium term this set-up might be employed to generate squeezed light, with the crucial advantage of exploiting the stronger nonlinear terms allowed by material degrees of freedom with respect to light, where nonlinear interactions giving rise to squeezing are always comparatively weak. More specifically, the squeezing of trapped ions could be previously obtained either by manipulating the trapping potentials \cite{heinzen90,noiartri,brown11} or through the internal degrees of freedom \cite{cirac93,meekhof96,solano05}, while in optomechanical systems, an high degree of squeezing could be obtained via indirect position measurements \cite{mskim-pnas}. 

The squeezing of a Gaussian state with CM $\sig$ has been quantified, in ${\rm dB}$, as $\max[0,-10\log_{10}(\sigma_{1}^{\uparrow})]$, where $\sigma_{1}^{\uparrow}$ is the smallest eigenvalue of $\sig$ (note that in our notation the vacuum state has eigenvalue $1$). 
In the following example, we have considered a trapped Indium ion, characterised by the experimental parameters shown in Table~\ref{table} (see also \ref{experimental-parameters}). Our quantitative findings are summarised in Fig.~\ref{figura}(a) showing that, even though the state transfer is not accurate for the considered experimental parameters (i.e., the curve differs from a straight line of unit slope), a substantial amount of squeezing can be transferred from the trapped ion to the output travelling mode. However, the output squeezing seems to hit a plateau at about $9$dB after an initial linear rise, indicating a complete breakdown of the approximated relation $f\simeq-ia_1(0)$ for high input squeezing. 

This is also reflected in the purity ${\rm Tr}(\varrho^{2})$ of the output state $\varrho$ which,  
being our state Gaussian, can be determined as $1/({\rm det}\vec \sigma_\textrm{\scriptsize out})^{1/2}$ \cite{serafozzi03}. 
Fig.~\ref{figura}(b) shows clearly that the purity of the output state is a decreasing function of the initial intra-cavity squeezing (in dB), 
proving that the conversion is not perfect since part of such squeezing is not coherently transferred but contributes instead to 
phase-insensitive noise in the output fields. 

\subsection{Entanglement generator.}
By considering, once again, output modes defined as in Eq.~\eqref{f}, we can also study the leakage of both cavity mirrors in a double-sided cavity, for 
an initial squeezed state of the mechanical oscillator, and consider the state of the two output modes leaking out of each mirror. Entanglement between those modes can be expected in this scenario, since the adiabatic elimination of the cavity field yields an effective coupling of the two output modes to the same motional degree of freedom. Indeed, it can be shown that the formal treatment of this case is identical to that of a single-sided cavity, up to mixing its output field with the vacuum at a beam-splitter (see \ref{two-sided}). For the Gaussian squeezed state of Section \ref{squeeze-transfer}, 
this operation results in entanglement between the two output modes, which can be quantified in terms of the logarithmic negativity $\N$ \cite{logneg}. 

For a two-mode Gaussian state with CM $\sig$, the logarithmic negativity can be evaluated as 
$\max[0,-\log_2(\tilde{\nu}_-)]$, where 
$\tilde{\nu}_{-}$ is the smallest eigenvalue of the matrix $|\Sigma T \sig T|$, with $T = \id_2\oplus\sigma_x$ ($\sigma_x$
being the $x$-Pauli matrix). 
Fig.~\ref{figura}(c) shows that a substantial amount of entanglement can be generated, and hence distributed,
for realistic amounts of internal squeezing. Our findings clearly point at the potential held by the motional degrees of freedom of massive particles as competitors of nonlinear crystals for the generation of optical entanglement.

\subsection{Continuous detection of the mechanical motion.}
Eq.~\eqref{boundary2} suggests that, by measuring the cavity output light with a fast detector, one should be able to monitor the mechanical motion in real time. In particular we shall focus here on the continuous detection of the mechanical population, keeping in mind that a similar analysis can be performed if one is interested in measuring the mechanical quadratures (see \ref{adiabatic}). We stress that here we are concerned with quantities averaged over many experimental runs, hence our treatment does not deal with the partial collapses of the quantum state that typically occur in a single realization of a continuous measurement \cite{wiseman}.

To formulate the problem in terms of an input-output  Gaussian channel, we consider modes of the form \eqref{g}, with $\mu_2=i\nu-\Gamma/2$, $\Gamma$ being the bandwidth of the detector. Again, the parameters $\mu_1,\mu_3$ can be set to any value, as we are only interested in the detection of the mode associated to $a_{\textrm{\scriptsize out},2}$, which we shall denote as $g(t)$. From Eqs.~\eqref{g} and \eqref{boundary2}, assuming that $\Gamma$ is large compared to the other frequencies in the problem, we get (see \ref{adiabatic})
\begin{equation}
	\langle{g(t)^\dagger g(t)}\rangle\simeq\frac{4\kappa}{\Gamma}\langle{a_1(t)^\dagger a_1(t)}\rangle.\label{temp-estimate}
\end{equation}
Therefore, by rescaling the average intensity registered by the detector, one should be able to infer the average number of phonons in the oscillator at time $t$.

Using our general formalism, we can now investigate quantitatively the validity of this prediction. In Fig.~\ref{figura}(d), we compare inferred and actual values of $\langle a_1^\dagger a_1\rangle$, as a function of time, for an Indium ion with an initial motional squeezing of $20{\rm dB}$. It can be seen that, after an initial transient time, the rescaled intensity at the detector follows faithfully and monotonically the mechanical population. Note however that the approximate relationship \eqref{temp-estimate} does not account for the detailed behaviour of the output field intensity, which is instead captured via our techniques.

\section{Conclusions}
The current developments in the control of harmonic oscillators at the micro-, nano- and atomic scale 
and in the technologies to couple them to travelling electromagnetic degrees of freedom \cite{optomechanics,ionlight}, as well as the wealth of applications quantum information theory has envisaged for such interfaces, 
call for compact and general frameworks to handle input-output processes in a variety of applied settings.
This work responds to such a need by delivering an algebraic description, in terms of Gaussian channels,
of input-output relationships for general white noise and quadratic interactions. 
As demonstrated, the applicability of the method is broad while its predictions are directly 
observable, and we hence believe it to hold potential for further applications in the context of 
input-output quantum interfaces.

\section*{Acknowledgments}
In fond memory of Wolfgang Lange. 

We thank V.Giovannetti, G. Milburn, A. Bayat, S. Bose, M. S. Kim and M. Tame for the useful discussions. We acknowledge financial support from
the Royal Society, Wolfson Foundation, UK EPSRC, Qatar National Research Fund
(NPRP 4-554-1-084), EU STREP HIP and PICC, EU Integrated
project QESSENCE, and the Alexander von Humboldt Foundation.

\section*{References}

\appendix
\section{Derivation of the Gaussian Channels}\label{channels}
\subsection{Exponential Pulses.} \label{channelA}
Let us start with the explicit derivation of the channel associated to Eq.~\eqref{f} . Combining with Eq.~\eqref{v}, 
and considering the $j$-th component, one has
\begin{eqnarray}
\hat f_j&=&|2{\rm Re}\lambda_j|^{1/2}\left[K \int_0^\infty{\rm d}t\;{\rm e}^{(A+\lambda_j\id)t} \hat{\vec{v}}(0) + K\int_0^\infty{\rm d}t\;{\rm e}^{(A+\lambda_j\id)t}  \int_{0}^{t} {\rm e}^{-As} \hat{\vec{v}}_\text{\scriptsize in}(s) {\rm d}s +\right.\nonumber\\ 
&&\left.\phantom{|2{\rm Re}\lambda_j|^{1/2}}- K^{-1}\int_0^\infty{\rm d}t\;{\rm e}^{\lambda_j t}\hat{\vec{v}}_\text{\scriptsize in}(t) \right]_j\label{fullfj}\\
	&=&-|2{\rm Re}\lambda_j|^{1/2}\left[K(A+\lambda_j\id)^{-1} \hat{\vec{v}}(0)+\right.\nonumber\\
	&&\phantom{-|2{\rm Re}\lambda_j|^{1/2}+}\left.+(K(A+\lambda_j\id)^{-1}+K^{-1})\int_0^\infty{\rm d}t\;{\rm e}^{\lambda_j t}\hat{\vec{v}}_\text{\scriptsize in}(t)\right]_j\;,\label{fjA}
\end{eqnarray}
where we have used that the primitive of a matrix exponential ${\rm e}^{Bt}$, in a domain where $B$ is invertible, is given by $B^{-1}{\rm e}^{Bt}$, and that all the exponentials involved in our calculations vanish in the limit $t\to\infty$. Integration by parts has been used to simplify the second term in Eq.~\eqref{fullfj}; more specifically, we have integrated the term ${\rm e}^{(A+\lambda_j\id)t}$ and differentiated the term $\int_{0}^{t} {\rm e}^{-As} \hat{\vec{v}}_\text{\scriptsize in}(s) {\rm d}s$. Eqs.~\eqref{fj}-\eqref{Y} follow by calculating explicitly the second moments of the modes $\hat f_j$. Combining Eq.~\eqref{fjA} with Eqs.~\eqref{white} and \eqref{uncorrelated}, and defining $\delta \hat f_j=\hat f_j-\av{\hat f_j}$ we have:
\begin{eqnarray}
	\langle\{\delta\hat{f}_j,\delta\hat{f}^{\dag}_k\}\rangle&=&|2{\rm Re}\lambda_j|^{1/2}|2{\rm Re}\lambda_k|^{1/2}\left[K(A+\lambda_j\id)^{-1}\vec{\sigma}(0)(K(A+\lambda_k\id)^{-1})^\dagger+\phantom{\frac{1}{2}}\right.\nonumber\\
	&&\left.-(K(A+\lambda_j\id)^{-1}+K^{-1})\frac{\vec{\sigma}_\textrm{\scriptsize in}}{\lambda_j+\lambda_k^*}(K(A+\lambda_k\id)^{-1}+K^{-1})^\dagger\right]_{jk}.
\end{eqnarray}
\subsection{Slow detectors and time-dependent spectral measurements}\label{channelB}
We now move on to the derivation of the Gaussian channel associated to Eq.~\eqref{g} . Combining Eqs.~\eqref{g} and \eqref{v} Considering the $j$-th component, we have

	\begin{eqnarray}
	\!\!\!\!	\hat g_j(t)&=&N_{t,j}\,{\rm e}^{\lambda_j t}\left[K\int_0^t{\rm d}t'\;{\rm e}^{(A-\lambda_j\id)t'}\hat{\vec{v}}(0)+K\int_0^t{\rm d}t'\;{\rm e}^{(A-\lambda_j\id)t'}\int_0^{t'}{\rm d}s\;{\rm e}^{-As}\hat{\vec{v}}_\textrm{\scriptsize in}(s)+\right.\nonumber\\
	&&\phantom{N_{t,j}\,{\rm e}^{\lambda_j t}}\left.-K^{-1}\int_0^t{\rm d}t'\;{\rm e}^{-\lambda_jt'}\hat{\vec v}_\textrm{\scriptsize in}(t')\right]_j\label{gA}\\
		&=&N_{t,j}\left[K(A-\lambda_j\id)^{-1}({\rm e}^{At}-{\rm e}^{\lambda_jt})\hat{\vec v}(0)+K(A-\lambda_j\id)^{-1}\int_0^t{\rm d}t'\;{\rm e}^{A(t-t')}\hat{\vec v}_\textrm{\scriptsize in}(t')+\right.\nonumber\\
	&&	\phantom{N_{t,j}}\left.-(K(A-\lambda_j\id)^{-1}+K^{-1})\int_0^t{\rm d}t'{\rm e}^{\lambda_j(t-t')}\hat{\vec v}_\textrm{\scriptsize in}(t')\right]_j,\label{gA1}
	\end{eqnarray}
where $N_{t,j}=\left|2{\rm Re}\lambda_j\right|^{1/2}[\id-{\rm e}^{2{\rm Re}(\lambda_j) t}]^{-1/2}$, and we have again performed integration by parts on the second term in Eq.~\eqref{gA}. The expression for the matrix $Z$ of Eq.~\eqref{Z}  follows easily from the first term in Eq.~\eqref{gA1}. To show how the matrix $T$ [Eq.~\eqref{T}] is derived, we shall now compute explicitly the second moments of the modes $\hat g_j$. Using Eqs.~\eqref{white} and \eqref{uncorrelated}, performing explicitly the integrals of the form $\int_0^t{\rm d}t'{\rm e}^{Bt'}=B^{-1}({\rm e}^{Bt}-\id)$, we have
	\begin{eqnarray}
		\langle\{ \delta \hat g_j,\delta \hat g_k^\dagger\}\rangle &&=[Z\vec\sigma(0)Z^\dagger]_{jk}+\nonumber\\
		&&+N_{t,j}N_{t,k}\left[K(A-\lambda_j\id)^{-1}\int_0^t{\rm d}t'{\rm e}^{At'}\vec\sigma_\textrm{\scriptsize in}{\rm e}^{A^\dagger t'}(K(A-\lambda_k\id)^{-1})^\dagger+\right.\nonumber\\
		&&+\left.\frac{{\rm e}^{(\lambda_j+\lambda_k^*)t}-1}{\lambda_j+\lambda_k^*}(K(A-\lambda_j\id)^{-1}+K^{-1})\vec{\sigma}_\textrm{\scriptsize in}(K(A-\lambda_k\id)^{-1}+K^{-1})^\dagger+\right.\nonumber\\
		&&\!\!\!\!\!\!\!\!\!\!\!\!\left.-K(A+\lambda_j\id)^{-1}(A+\lambda_k^*\id)^{-1}({\rm e}^{(A+\lambda_k^*\id)t}-\id)\vec\sigma_\textrm{\scriptsize in}(K(A-\lambda_k\id)^{-1}+K^{-1})^\dagger+\right.\nonumber\\
		&&\!\!\!\!\!\!\!\!\!\!\!\!\!\!\!\!\!\!\left.-(K(A-\lambda_j\id)^{-1}+K^{-1})\vec\sigma_\textrm{\scriptsize in}(A^\dagger+\lambda_j\id)^{-1}({\rm e}^{(A^\dagger+\lambda_j\id)t}-\id)(K(A-\lambda_k\id)^{-1})^\dagger\right]_{jk}.\nonumber\\\label{channel2}
	\end{eqnarray}
From Eq.~\eqref{channel2}, one can derive easily Eqs.~\eqref{T}-\eqref{I}. The only integral which cannot be performed straightforwardly is given by
\begin{equation}
	I=\int_0^t{\rm d}t'{\rm e}^{At'}\vec\sigma_\textrm{\scriptsize in}{\rm e}^{A^\dagger t'}.\label{integral}
\end{equation}
As emphasized in the main text, one can verify explicitly that $I$ verifies a Sylvester equation
\begin{equation}
	AI+IA^\dagger=\int_0^t{\rm d}t'\;\frac{{\rm d}}{{\rm d}t'}\left({\rm e}^{At'}\vec\sigma_\textrm{\scriptsize in}{\rm e}^{A^\dagger t'}\right)={\rm e}^{At}\vec{\sigma}_\text{\scriptsize in}{\rm e}^{A^{\dag}t}-\vec{\sigma}_\text{\scriptsize in}.\label{sylvestro}
\end{equation}
It is known that Eq.~\eqref{sylvestro} has a unique solution if and only if $A$ and $-A^\dagger$ have no common eigenvalues \cite{Sylvester}. Note that in our system this condition is automatically satisfied. In fact, the hypothesis of stability for the matrix $A$ (that is, $\lim_{t\to\infty}{\rm e}^{At}=0$), implies that the eigenvalues of $A$ have strictly negative real parts, hence those of $-A^\dagger$ must have positive real parts. Consequently, the unique algebraic solution to Eq.~\eqref{sylvestro} provides the result of the integral in Eq.~\eqref{integral}.
\section{Ion-cavity Hamiltonian}\label{ion-cavity}
Our system is composed of a two-level ion, trapped inside an optical cavity and driven by an external laser beam. The internal levels of the ion are denoted as $\ket{e}, \ket{g}$, have frequency splitting $\omega_0$ and their dynamics is conveniently described via a set of Pauli operators $\sigma_z,\sigma^\pm$ (In our notation, we have $\sigma_z=\kebra{e}{e}-\kebra{g}{g},\sigma^+=\kebra{e}{g},\sigma^-=\kebra{g}{e}$). We assume that it is sufficient to consider a single mode of the electromagnetic field inside the optical cavity, with annihilation operator $a_2$ and frequency $\omega_c$. We also assume that the motion of the center of mass of the ion is relevant only along one axis, which we may choose as $x$ without loss of generality. The total Hamiltonian, in a frame rotating with the frequency $\omega_\textrm{\scriptsize L}$ of the driving laser, reads
\begin{eqnarray}
	\! \! \! \! \! \! \! \! \! \! \! \! \! \! \! \! H_\textrm{\scriptsize tot}=\nu a_1^\dagger a_1+\delta a_2^\dagger a_2+\frac{\Delta}{2}\sigma_z+g_0\left(a_2^\dagger \sigma^-+a_2\sigma^+\right)+i\Omega\left(\sigma^+e^{ikx}-\sigma^-e^{-ikx}\right)
\end{eqnarray}
where $a_1$ is the annihilation operator for the ion motion along the $x$ axis [$x=x_0(a_1+a_1^\dagger)$, where $x_0$ is the ground state spread], $\nu$ the motional trapping frequency, ${\delta=\omega_c-\omega_\textrm{\scriptsize L}}$, ${\Delta=\omega_0-\omega_\textrm{\scriptsize L}}$ are respectively the detuning of the cavity and the ion internal transition with respect to the laser frequency, $g_0$ is the strength of the Jaynes-Cummings interaction between the internal levels of the ion and the cavity field, while $\Omega$ is the Rabi frequency of the driving laser, $k$ being the projection of the light wavevector on the $x$ axis ($k\leq\omega_\textrm{\scriptsize L}/c$). 
Note that, as a consequence of its motion, the ion experiences variations in the phase of the driving field, while the strength of the Jaynes-Cummings interaction remains approximately constant. This is the case if, for example, the equilibrium position of the ion coincides with a maximum of the cavity field intensity. In the Lamb-Dicke regime $k\sqrt{\av{x^2}}\ll1$, we can approximate ${\rm e}^{ikx}\simeq1+ikx$, hence
\begin{eqnarray}
\!\!\!\!\!\!\!\!\!\!\!\!\!\!\!\!\!\! H_\textrm{\scriptsize tot}&&\simeq \nu a_1^\dagger a_1+\delta a_2^\dagger a_2+\frac{\Delta}{2}\sigma_z+g_0\left(a_2^\dagger \sigma^-+a_2\sigma^+\right)+i\Omega\left(\sigma^+-\sigma^-\right)+\nonumber\\
&&-\eta\Omega\left(\sigma^++\sigma^-\right)\left(a_1+a_1^\dagger\right),
\end{eqnarray}
where $\eta=kx_0$ is the Lamb-Dicke Parameter. As we have emphasized in the main text, we are interested in a large detuning regime in which the internal transition of the ion is only virtually excited (that is, $\Delta\gg g_0,\Omega,\nu,\delta$). Therefore, we shall introduce only a small error by substituting $\sigma_z\simeq 2a_3^\dagger a_3-1,\sigma^+\simeq a_3^\dagger,\sigma^-\simeq a_3$, where $a_3$ is a bosonic mode. In this way, the problem becomes tractable in our general language of Input-output Gaussian channels. The resulting Hamiltonian, up to a constant, is then given by
\begin{eqnarray}
\!\!\!\!\!\!\!\!\!\!\!\!\!\!\!\!\!\!	H_\textrm{\scriptsize tot}&&\simeq \nu a_1^\dagger a_1+\delta a_2^\dagger a_2+\Delta a_3^\dagger a_3+g_0\left(a_2^\dagger a_3+a_2a_3^\dagger\right)+i\Omega\left(a_3-a_3^\dagger\right)+\nonumber\\
&&-\eta\Omega\left(a_3+a_3^\dagger\right)\left(a_1+a_1^\dagger\right).\label{Hion}
\end{eqnarray}
Now, the quadratic part of the Hamiltonian in Eq.~\eqref{Hion} coincides with the expression found in Eq.~\eqref{coupling2} , while the linear part can be included in the equations of motion by adding a constant to the vector $\bar{\vec{v}}_\text{\scriptsize in}$ (see main text). In order for the bosonic treatment of the two-level system to be consistent, one has to check a posteriori that, at all times, the population of the mode $a_3$ is confined to the ground and first excited state with good approximation. Below we show estimates for the probability that the mode $a_3$ occupies excited states above the first [indicated as $P(a_3^\dagger a_3>1)$], as a function of rescaled time, for the applications described in the main text (the parameters used are given in Table~\ref{table}). We have fixed the initial motional squeezing to $20 {\rm dB}$, which yields the largest occupation for the mode $a_3$ in the considered parameter range. Note how the results are consistent with the bosonic approximation of the two level system.
\begin{center}
	\includegraphics[width=.48\textwidth]{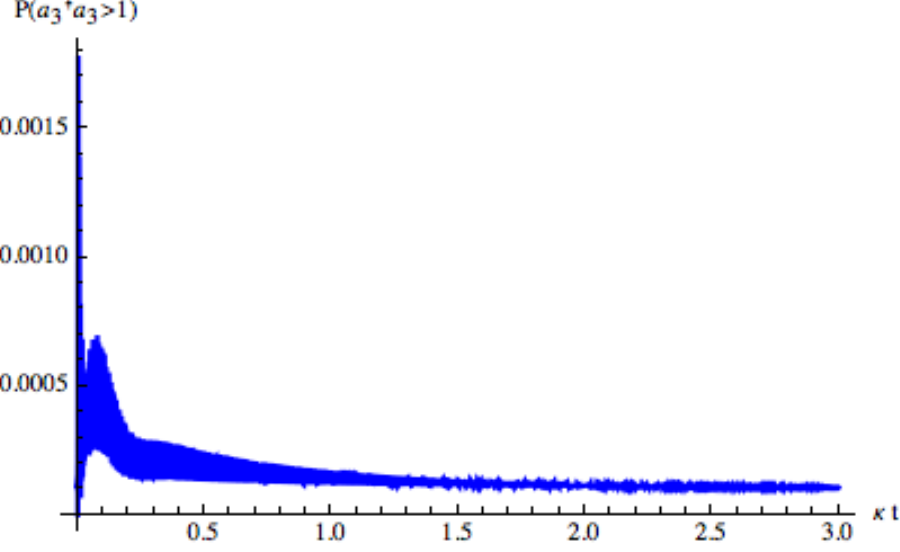}
\end{center}
This quantity has been calculated by explicitly integrating Eq.~\eqref{langio}, using the full Hamiltonian of Eq.~\eqref{Hion}, and assuming a thermal distribution for the population of the mode $a_3$, so that the only relevant parameter is $\langle a_3^\dagger a_3\rangle$, which can calculated via standard input-output theory. One then has the estimate
\begin{equation}
	P(a_3^\dagger a_3>1)\sim\left(\frac{\langle a_3^\dagger a_3\rangle}{1+\langle a_3^\dagger a_3\rangle}\right)^2.
\end{equation}

\section{Adiabatic elimination of the cavity field}\label{adiabatic}
Let us now take Eq.~\eqref{coupling}  as a starting point. Taking $\delta'=\nu$, and assuming $J\ll\nu$, we can use a Rotating Wave Approximation \cite{james10,james11} to neglect counter-rotating terms. Hence, we have
\begin{equation}
	\hat{H}' \simeq \nu a_1^{\dag}a_1 + \nu a_2^{\dag}a_2 + J(a_1a_2^\dagger+a_2a_1^{\dag}) \;.
\end{equation}
Correspondingly, the equations of motion for the bosonic modes read
\begin{eqnarray}
	\dot a_1&=-i\nu a_1-\frac{\kappa_1}{2}a_1-iJa_2+\sqrt{\kappa_1}a_\textrm{\scriptsize in,1},\label{ad1}\\
	\dot a_2&=-i\nu a_2-\frac{\kappa_2}{2}a_2-iJa_1+\sqrt{\kappa_2}a_\textrm{\scriptsize in,2}.\label{ad2}
\end{eqnarray}
If the cavity is initially empty, and the coupling of the cavity field to the mechanical mode is weak, i.e. $J\ll \kappa_2$, we can expect that the mode $a_2$ will remain in its ground state with good approximation. In the Heisenberg picture, this amounts to taking $\dot a_2\simeq-i\nu a_2$. Substituting in Eq.~\eqref{ad2}, we get
\begin{equation}
	a_2\simeq-i\frac{2J}{\kappa_2}a_1+\frac{2}{\sqrt{\kappa_2}}a_\textrm{\scriptsize in,2}.\label{ad3}
\end{equation}
Defining $\kappa=4J^2/\kappa_2$, substituting Eq.~\eqref{ad3} in Eq.~\eqref{ad1} and considering the limit $\kappa\gg\bar n_\textrm{\scriptsize th}\kappa_1$ (so that the environment associated to $a_\textrm{\scriptsize in,1}$ can be neglected), we get Eq.~\eqref{langio2}. To get Eq.~\eqref{boundary2}, one simply has to combine Eq.~\eqref{ad3} with the boundary condition $a_\textrm{\scriptsize out,2}=\sqrt{\kappa_2}a_2-a_\textrm{\scriptsize in,2}$.

Let us now show how to obtain Eq.~\eqref{temp-estimate}. The bosonic mode of interest (see main text) has the explicit form
\begin{equation}
	g(t)\simeq\frac{\Gamma^{1/2}}{(1-{\rm e}^{-\Gamma t})}\int_0^t{\rm d}t'\;{\rm e}^{(i\nu-\Gamma/2)(t-t')}(-i\sqrt{\kappa}a_1(t')-a_\textrm{\scriptsize in,2}(t')),
\end{equation}
where we have already taken into account the adiabatic elimination of the mode $a_2$. In the limit where $\Gamma$ is large compared to any other frequency in the problem, me may assume that the main contribution to the integral is achieved for $t'\sim t$. Hence, we may replace $a_1(t')\simeq a_1(t)$ and perform the integration explicitly, yielding
\begin{equation}
	g(t)\simeq2i\sqrt{\frac{\kappa}{\Gamma}}a_1(t)+\xi_\textrm{\scriptsize in},\label{timecool}
\end{equation}
where $\xi_\textrm{\scriptsize in}$ is a noise operator resulting from the integration of $a_\textrm{\scriptsize in,2}$, and we have approximated ${\rm e}^{-\Gamma/2 t}\simeq0$. By construction, the bosonic mode associated to $\xi_\textrm{\scriptsize in}$ is in the vacuum state. Therefore, Eq.~\eqref{temp-estimate} can be derived easily from Eq.~\eqref{timecool}. In addition, one has
\begin{equation}
\langle g(t)^2\rangle\simeq\frac{4\kappa}{\Gamma}\langle a_1(t)^2\rangle.\label{g^2}
\end{equation}
Combining appropriately Eqs.~\eqref{temp-estimate} and \eqref{g^2}, the relationship between the statistics of arbitrary quadratures of $g(t)$ and $a_1(t)$ can be derived.
\section{Two-sided cavities, and losses in the output fields}\label{two-sided}
Consider a cavity mode $a$, such that photons can leak out of both mirrors (left or right) with rates $\kappa_\textrm{\scriptsize L},\kappa_\textrm{\scriptsize R}$ respectively. In addition, the mode is subject to some Hamiltonian $H$. The corresponding Quantum Langevin equation reads
\begin{equation}
	\dot a=i[H,a]-\frac{\kappa_\textrm{\scriptsize L}+\kappa_\textrm{\scriptsize R}}{2}a+\sqrt{\kappa_\textrm{\scriptsize L}}a_\textrm{\scriptsize in,\scriptsize{L}}+\sqrt{\kappa_\textrm{\scriptsize R}}a_\textrm{\scriptsize in,\scriptsize{R}},
\end{equation}
where $a_\textrm{\scriptsize in,\scriptsize{L}},a_\textrm{\scriptsize in,\scriptsize{R}}$ are two uncorrelated and independent bosonic noise operators, with the usual properties. The output fields on the two sides of the cavity are defined via the boundary conditions:
\begin{eqnarray}
	&a_\textrm{\scriptsize out,L}=\sqrt{\kappa_\textrm{\scriptsize L}}a-a_\textrm{\scriptsize in,\scriptsize{L}},\\
	&a_\textrm{\scriptsize out,R}=\sqrt{\kappa_\textrm{\scriptsize R}}a-a_\textrm{\scriptsize in,\scriptsize{R}}.
\end{eqnarray}
Now, let us consider the beam-splitter transformation
\begin{eqnarray}
	BS=\frac{1}{\sqrt{\kappa_\textrm{\scriptsize tot}}}\left(\begin{array}{cc}
	\sqrt{\kappa_\textrm{\scriptsize L}}&\sqrt{\kappa_\textrm{\scriptsize R}}\\
	-\sqrt{\kappa_\textrm{\scriptsize R}}&\sqrt{\kappa_\textrm{\scriptsize L}}
	\end{array}\right),
\end{eqnarray}
where we have defined $\kappa_\textrm{\scriptsize tot}=\kappa_\textrm{\scriptsize L}+\kappa_\textrm{\scriptsize R}$. We can now define the transformed input-output operators
\begin{eqnarray}
	&\left(\begin{array}{cc}
		a_\textrm{\scriptsize in}\\
		\xi_\textrm{\scriptsize in}
	\end{array}\right)=BS\left(\begin{array}{cc}
		a_\textrm{\scriptsize in,\scriptsize{L}}\\
		a_\textrm{\scriptsize in,\scriptsize{R}}
	\end{array}\right),\\
	&\left(\begin{array}{cc}
		a_\textrm{\scriptsize out}\\
		\xi_\textrm{\scriptsize out}
	\end{array}\right)=BS\left(\begin{array}{cc}
		a_\textrm{\scriptsize out,L}\\
		a_\textrm{\scriptsize out,R}
	\end{array}\right).\label{BSout}
\end{eqnarray}
In terms of these, the equation of motion for $a$ and the boundary conditions for the input-output fields can be rewritten as
\begin{eqnarray}
	&\dot a=i[H,a]-\frac{\kappa_\textrm{\scriptsize tot}}{2}a+\sqrt{\kappa_\textrm{\scriptsize tot}}a_\textrm{\scriptsize in},\\
	&a_\textrm{\scriptsize out}=\sqrt{\kappa_\textrm{\scriptsize tot}}a-a_\textrm{\scriptsize in},\\
	&\xi_\textrm{\scriptsize out}=-\xi_\textrm{\scriptsize in},
\end{eqnarray}
where we see that the cavity mode now couples only to the noise operator $a_\textrm{\scriptsize in}$, while the mode $\xi_\textrm{\scriptsize in}$ is simply reflected. Our general theory as described in the main text can be applied to the effective output mode $a_\textrm{\scriptsize out}$, however in this case the physically meaningful modes are associated to the outputs of the left and right mirror, according to a relation of the type
\begin{eqnarray}
	\left(\begin{array}{cc}
f_\textrm{\scriptsize L}\\f_\textrm{\scriptsize R}
	\end{array}\right)=\int{\rm d}t\; \vartheta(t)\left(\begin{array}{cc}
		a_\textrm{\scriptsize out,L}(t)\\
		a_\textrm{\scriptsize out,R}(t)
	\end{array}\right),\label{fLfR}
\end{eqnarray}
where for simplicity we have assumed the same temporal profile $\vartheta(t)$. Different temporal profiles can also be accounted for, but they are outside the scopes of this paper. Let us now show the relation between the physical modes $f_\textrm{\scriptsize L},f_\textrm{\scriptsize R}$, and the mode that can be calculated with our theory. By inverting Eq.~\eqref{BSout}, and substituting in Eq.~\eqref{fLfR}, one can easily see that
\begin{eqnarray}
	&\left(\begin{array}{cc}
	f_\textrm{\scriptsize L}\\f_\textrm{\scriptsize R}
	\end{array}\right)=BS^\dagger \left(\begin{array}{cc}
	f\\ \xi
	\end{array}\right),\\
	&\left(\begin{array}{cc}
	f\\ \xi
	\end{array}\right)=\int{\rm d}t\; \vartheta(t)
	\left(\begin{array}{cc}
		a_\textrm{\scriptsize out}(t)\\
		\xi_\textrm{\scriptsize out}(t)
	\end{array}\right),
\end{eqnarray}
where we see explicitly that a two-sided cavity is equivalent to a one-sided cavity followed by a beam-splitter (with the parameters given above). Note that the statistics of $f$ can be calculated via the methods given in the main text, while that of $\xi$ is uniquely determined by the noise operators $a_\textrm{\scriptsize in,\scriptsize{L}},a_\textrm{\scriptsize in,\scriptsize{R}}$ (Typically, $\xi$ will be in the vacuum or in a thermal state).

As a final remark, we note that the mathematics of a two-sided cavity described here can be also used to describe partial losses in the output field. Indeed, one can take $a_\textrm{\scriptsize out,L}$ to describe the accessible portion of the output field, while $a_\textrm{\scriptsize out,R}$ can be interpreted as the portion that is lost (e.g. due to scattering and absorption on the mirrors). Then, only $f_L$ describes the physically accessible bosonic mode, while the mode $f_R$ can be traced out.
\section{Experimental parameters}\label{experimental-parameters}
The parameters of Table~\ref{table} can be obtained by considering the $\lambda=230{\rm nm}$ transition of an Indium ion \cite{indium}. For this transition, the decay rate is $\kappa_3=2\pi\cdot360{\rm kHz}$. Assuming that the ion equilibrium position coincides with a maximum of the cavity field, the Jaynes-Cummings coupling strength is given by \cite{kimble}
\begin{equation}
	g_0=\frac{1}{2}\sqrt{\frac{3\lambda^2 c\kappa_3}{2\pi V}},
\end{equation}
where $c$ is the speed of light in vacuum, and $V$ is the effective volume of the cavity field. For a Gaussian ${\rm TEM_{00}}$ mode, we have \cite{kimble}
\begin{equation}
	V\simeq L\times\pi\left(\frac{w_0}{2}\right)^2,
\end{equation}
where $L$ is the length of the cavity and $w_0$ the waist of the mode at the cavity center. Assuming $L=1cm$ and $w_0=6{\rm \mu m}$ we get
\begin{equation}
	g_0\simeq2\pi\cdot0.62{\rm MHz}.
\end{equation}
Let us briefly discuss the remaining parameters. The value $\kappa_2=2\pi\cdot53{\rm kHz}$, as given in the main text, corresponds to a cavity finesse
\begin{equation}
	\mathcal F=\frac{\pi c}{L\kappa_2}\simeq 2.8\cdot 10^5.
\end{equation} 
A motional trap frequency of $\nu=2\pi\cdot5{\rm MHz}$, combined with a mass of $m\simeq114{\rm a.u.}$ for the Indium ion, yields the Lamb-Dicke parameter (in our units, $\hbar=1$)
\begin{equation}
	\eta=\frac{2\pi x_0}{\lambda}=\frac{2\pi}{\lambda}\sqrt{\frac{1}{2m\nu}}\simeq0.081.
\end{equation}
Finally, we assume that the ion trap is stabilized to provide an average heating time $(\bar n_\textrm{\scriptsize th}\kappa_1)^{-1}=6.6{\rm ms}$ at $T=300$K. This implies an average phonon number
\begin{equation}
	\bar n_\textrm{\scriptsize th}\simeq1.2\cdot10^6,
\end{equation}
and an heating rate per phonon
\begin{equation}
	\kappa_1\simeq2\pi\cdot2\cdot10^{-5}{\rm Hz}.
\end{equation}

\begin{thebibliography}{99}

\bibitem{divincenzo} See the ARDA {\em Quantum Information Science and Technology Roadmap} at http://qist.lanl.gov.

\bibitem{parkins99} A. S. Parkins and H. J. Kimble, J. Opt. B: Quantum Semiclass. Opt. {\bf 1}, 496 (1999).

\bibitem{parkins00} A. S. Parkins and E. Larsabal, Phys. Rev. A {\bf 63}, 012304 (2000). 

\bibitem{Zhang03} J. Zhang, K. Peng, and S. L. Braunstein, Phys. Rev. A {\bf 68}, 013808 (2003). 

\bibitem{morigi06} G. Morigi, J. Eschner, S. Mancini, and D. Vitali, Phys. Rev. Lett. {\bf 96}, 023601 (2006). 

\bibitem{kimble08} H. J. Kimble, Nature {\bf 453}, 1023 (2008).

\bibitem{martin} S. Bose, P. L. Knight, M. B. Plenio and V. Vedral, Phys. Rev. Lett. \textbf{83}, 5158 (1999); X-L Feng et al, Phys. Rev. Lett. \textbf{90}, 217902 (2003); D. E. Browne, M. B. Plenio, S. F. Huelga, Phys. Rev. Lett. \textbf{91}, 067901 (2003).

\bibitem{martin2} M. B. Plenio et al, Phys. Rev. A \textbf{59}, 2468 (1999); A. S. Sorensen, K. Molmer, Phys. Rev. Lett. \textbf{90}, 127903 (2003); A. S. Sorensen, K. Molmer,
Phys. Rev. Lett. \textbf{91}, 097905 (2003); L.-M. Duan, H. J. Kimble, Phys. Rev. Lett. \textbf{90}, 253601 (2003). 

\bibitem{yurke84} B. Yurke, Phys. Rev. A {\bf 29}, 408 (1984).

\bibitem{collett84} M. J. Collett and C. Gardiner, Phys. Rev. A {\bf 30}, 1386 (1984).

\bibitem{wallsmilburn} D. F. Walls and G. J. Milburn, {\em Quantum Optics} (Springer, Berlin, 2008).

\bibitem{demoen77} B. Demoen, P. Vanheuswijn, and A. Verbeure, Lett. Math. Phys. {\bf 2}, 161 (1977). 

\bibitem{eisertplenio} J. Eisert and  M. B. Plenio, Int. J. Quant. Inf. {\bf 1}, 479 (2003).

\bibitem{eisert07} J. Eisert and M. M. Wolf, {\em Gaussian quantum channels}, in {\em Quantum Information with Continous Variables of Atoms and Light}, N. J. Cerf, G. Leuchs, and E. S. Polzik Eds., pp. 23-42 (Imperial College Press, London, 2007).

\bibitem{eberly} J. H. Eberly and K. W\'{o}dkiewicz, J. Opt. Soc. Am. \textbf{67}, 1252 (1977) 

\bibitem{Sylvester} R. H. Bartels and G. W. Stewart, Comm. ACM, \textbf{15}, 820 (1972).

\bibitem{optomechanics} F. Marquardt and S. M. Girvin, Physics \textbf{2}, 40 (2009); T. J. Kippenberg and K. J. Vahala, Science \textbf{321}, 1172 (2008).

\bibitem{zippilli05} S. Zippilli and G. Morigi, Phys. Rev. Lett. {\bf 95}, 143001 (2005).

\bibitem{machnes12} S. Machnes, J. Cerrillo, M. Aspelmeyer, W. Wieczorek, M. B. Plenio, and A. Retzker, Phys. Rev. Lett. {\bf 108}, 153601 (2012).

\bibitem{wilson} I. Wilson-Rae, N. Nooshi, W. Zwerger, and T. J. Kippenberg,
Phys. Rev. Lett. \textbf{99}, 093901 (2007); Florian Marquardt, Joe P. Chen, A. A. Clerk, and S. M. Girvin, Phys. Rev. Lett. \textbf{99}, 093902 (2007).

\bibitem{james10} O. Gamel and D.F.V. James, Phys. Rev. A {\bf 82}, 052106 (2010).

\bibitem{james11} D.F.V. James and O. Gamel, Can. J. Phys. {\bf 85}, 625 (2011).

\bibitem{heinzen90} D. J. Heinzen and D. J. Wineland, Phys. Rev. A {\bf 42}, 2977 (1990). 

\bibitem{noiartri} A. Serafini, A. Retzker, and M.B. Plenio, New. J. Phys. {\bf 11}, 023007 (2009);
Quantum Inf. Process. {\bf 8}, 619 (2009).

\bibitem{james-ions} D. F. V. James, Appl. Phys. B \textbf{66}, 181 (1998).


\bibitem{wiseman} H. M. Wiseman and G. J. Milburn, \textit{Quantum Measurement and Control}, Cambridge University Press, Cambridge (2009).

\bibitem{indium} J. von Zanthier, M. Eichenseer, A. Yu. Nevsky, M. Okhapkin,
Ch. Schwedes and H. Walther, Laser Phys. \textbf{7}, 1021 (2005).

\bibitem{kimble} H. J. Kimble, in \textit{Cavity Quantum Electrodynamics}, edited by P. R. Berman, (Academic, New York, 1994) p. 203.

\bibitem{brown11} K.R. Brown {\em et al.}, Nature {\bf 471}, 196 (2011).

\bibitem{cirac93} J.I. Cirac, A.S. Parkins, R. Blatt, and P. Zoller, Phys. Rev. Lett. {\bf 70}, 556 (1993).

\bibitem{meekhof96}{D.M. Meekhof {\em et al.}, Phys. Rev. Lett. {\bf 76}, 1796 (1996).}

\bibitem{solano05} E. Solano, Phys. Rev. A {\bf 71}, 013813 (2005).

\bibitem{mskim-pnas} M. R. Vanner, I. Pikovski, G. D. Cole, M. S. Kim, C. Brukner, K. Hammerer, G. J. Milburn, M. Aspelmeyer, Proc. Natl. Acad. Sci. USA \textbf{108}, 16182 (2011).

\bibitem{serafozzi03} M. G. A. Paris, F. Illuminati, A. Serafini, and S. De Siena, Phys. Rev. A {\bf 68}, 012314 (2003); G. Adesso, A. Serafini, F. Illuminati, Phys. Rev. Lett. \textbf{92}, 087901 (2004).

\bibitem{logneg} J. Eisert and M.B. Plenio, J. Mod. Opt. {\bf 46}, 145 (1999); 
J. Lee, M.S. Kim, Y.J. Park, and S. Lee, J. Mod. Opt. {\bf 47}, 2151 (1999); 
G. Vidal and R.F. Werner, Phys. Rev. A {\bf 65}, 032314 (2002);
M.B. Plenio Phys. Rev. Lett. {\bf 95}, 090503 (2005).

\bibitem{ionlight}{G. R. Guth{\"o}hrlein, M. Keller, K. Hayasaka, W. Lange, and H. Walther, Nature {\bf 414}, 49 (2001);
M. Keller, B. Lange, K. Hayasaka, W. Lange, and H. Walther, Nature {\bf 431}, 1075 (2004); 
P. F. Herskind, A. Dantan, J. P. Marler, M. Albert, and M. Drewsen, Nature Phys. {\bf 5}, 494 (2009).}

\end{thebibliography}
\end{document}